\documentclass[twocolumn,showpacs,preprintnumbers,amsmath,amssymb,prx]{revtex4-1}
\usepackage{graphicx}
\usepackage{color}
\usepackage{amsmath,amsthm,amssymb}
\usepackage{braket}
\begin{document}
 
\title{First-Principles Calculation of Spin and Orbital Contributions to Magnetically Ordered Moments in Sr$_{2}$IrO$_{4}$}

\author{Christopher Lane*\textsuperscript{1}}
\author{Yubo Zhang\textsuperscript{2}}
\author{James W. Furness\textsuperscript{2}}
\author{Robert S. Markiewicz\textsuperscript{1}}
\author{Bernardo Barbiellini\textsuperscript{3,1}}
\author{Jianwei Sun*\textsuperscript{2}}
\author{Arun Bansil*\textsuperscript{1}}

\affiliation{
\textsuperscript{1}Department of Physics, Northeastern University, Boston MA 02115, USA \\
\textsuperscript{2}Department of Physics and Engineering Physics, Tulane University, New Orleans, LA 70118, USA\\
\textsuperscript{3}Department of Physics, School of Engineering Science, LUT University, FI-53851 Lappeenranta, Finland\\
}

\date{version of \today} 
\begin{abstract}
We show how an accurate first-principles treatment of the canted-antiferromagnetic ground state of Sr$_2$IrO$_4$, a prototypical $5d$ correlated spin-orbit coupled material, can be obtained without invoking any free parameters such as the Hubbard U or tuning the spin-orbit coupling strength. Our theoretically predicted iridium magnetic moment of 0.250 $\mu_B$, canted by 12.6$^{\circ}$ off the a-axis, is in accord with experimental results. By resolving the magnetic moments into their spin and orbital components, we show that our theoretically obtained variation of the magnetic scattering amplitude $\braket{M_{m}}$ as a function of the polarization angle is consistent with recent non-resonant magnetic x-ray scattering measurements. The computed value of the band gap (55 meV) is also in line with the corresponding experimental values. A comparison of the band structure to that of the cuprates suggests the presence of incommensurate charge-density wave phases in Sr$_{2}$IrO$_{4}$.
\end{abstract}


\maketitle 
\section{Introduction}
Over the past decade, it has become clear that spin-orbit coupling (SOC) is a key player in driving exotic physics in quantum matter. For example, spin-orbit coupling can modify electronic band structures to produce a variety of topological insulators and semimetals\cite{Bansil2016}. Spin-orbit coupling influences the magnetic exchange coupling to generate phase diagrams that include spin-liquids and charge fractionalization\cite{Jackeli2009,Hermanns2018}. It also plays an important role in the physics of heavy fermion systems and their unusual non-Fermi liquid behavior and unconventional superconductivity\cite{Stewart1979,Stockert2011,Pfleiderer2009}. 

5$d$ transition-metal oxides are interesting in this connection since they involve interplay of electron-electron interactions and strong spin-orbit coupling effects. In particular, the Ruddlesden-Popper single layer iridate, Sr$_{2}$IrO$_{4}$, has gained substantial attention for its striking similarity to La$_2$CuO$_4$ (LCO), a prototypical cuprate high-temperature superconductor. In Sr$_{2}$IrO$_{4}$, the Ir$^{4+}$ 5$d$ $t_{2g}$ states are split by spin-orbit coupling to produce a half-filled $J_{eff}=1/2$ band much like the half-filled CuO$_2$ band in LCO.\cite{Kim2008} The $J_{eff}=1/2$ states have a reduced bandwidth, such that a moderate onsite Hubbard potential $U$ is sufficient to drive the system toward an antiferromagnetic (AFM) instability. Experimentally, Sr$_2$IrO$_4$ is found to be an AFM insulator in which Fermi arcs have been reported upon electron doping\cite{Kim2014} and a low-energy nodal kink\cite{hu2019spectroscopic}. However, despite the similarity to LCO, superconductivity has not been reported.

Recently, the validity of the $J_{eff}=1/2$ description has come into question, igniting a debate over the exact nature of the ground state of Sr$_2$IrO$_4$.\cite{sala2014resonant} Polarized neutron diffraction measurements show an anisotropic (aspherical) magnetization density distribution of primarily Ir $d_{xy}$ character. \cite{jeong2019magnetization}  
Resonant inelastic x-ray scattering finds strong hybridization between the  IrO$_6$ octahedra due to the delocalized 5$d$ orbitals, complicating the strictly local picture of the low-energy electronic structure.\cite{Agrestini2017,Wang2011} Additionally, non-resonant magnetic x-ray scattering (NRMXS) finds the branching ratio $\braket{L}/\braket{S}$ deviating from the $J_{eff}=1/2$ model.\cite{Fujiyama2014} Interestingly, intertwining of the nearly degenerate low-energy magnetic groundstates with the lattice degrees-of-freedom\cite{DiMatteo2016,porras2019pseudospin}, similar to the case of the yttrium-based cuprates\cite{zhang2018landscape,markiewicz2019first} might also be at play. 

\begin{figure*}[t]
\includegraphics[width=.99\textwidth]{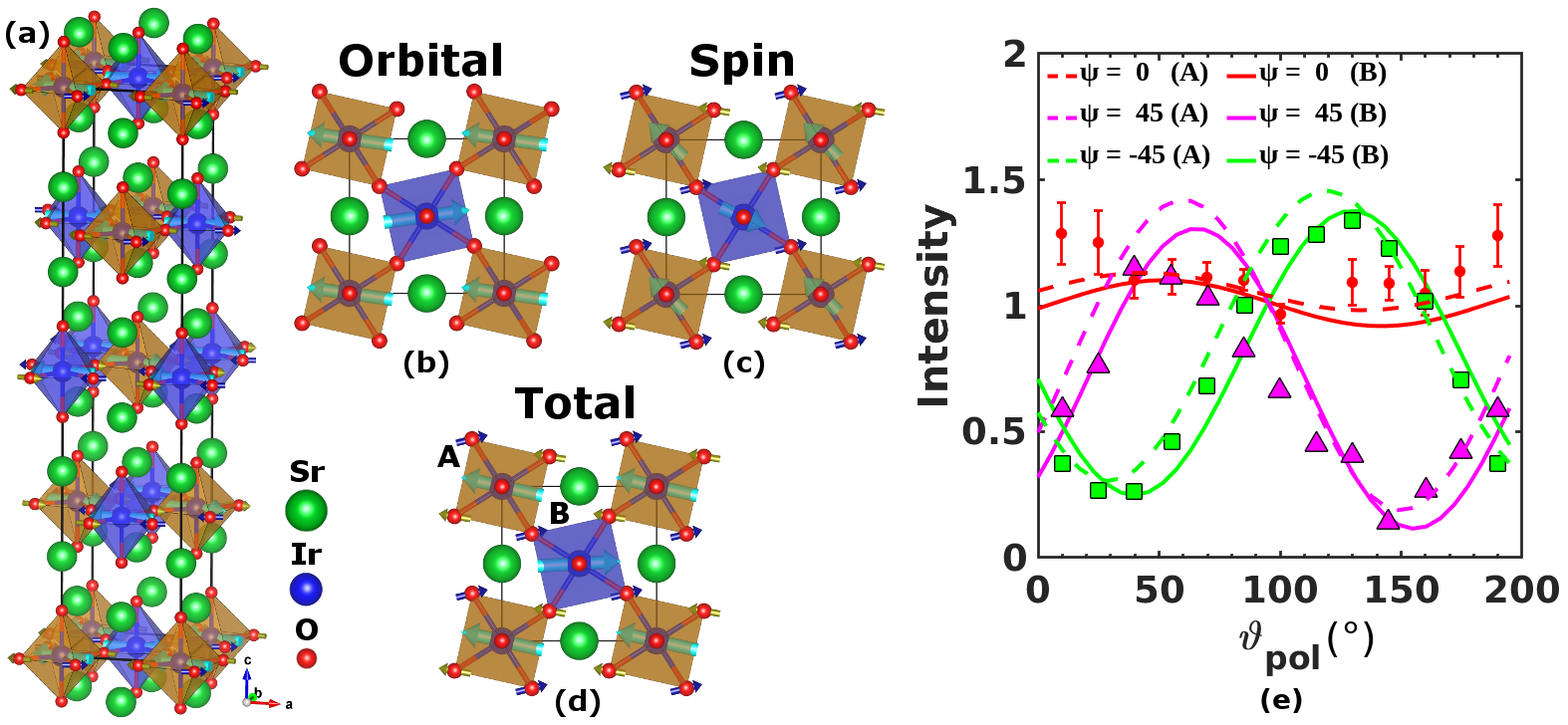}
\caption{(color online) (a) Theoretically predicted AFM state of Sr$_2$IrO$_4$ where the iridium, oxygen and strontium atoms are represented by blue, red and green spheres, respectively. Light-blue arrows represent Ir magnetic moments. The in-plane oxygen atoms have a slight net magnetic moment (blue and gold arrows represent inequivalent moments). Staggered octahedral rotations are highlighted by coloring the octahedra dark orange (blue) for clockwise (counter-clockwise) twists; black lines mark the unit cell. Panels (b) and (c) show the orbital and spin components of the magnetic moment in a single IrO$_{2}$ layer, whereas the total moment is shown in panel (d). (e) Theoretical (solid and dashed lines) and experimental \cite{Fujiyama2014} (symbols) $\vartheta_{\text{pol}}$ dependence of NRMXS amplitude for three azimuthal angles ($\Psi$) for Ir sites $A$ and $B$ in the AFM ground state of Sr$_{2}$IrO$_{4}$.}
\label{fig:SPINSTRUCTURE}
\end{figure*}

An accurate first-principles treatment of correlated materials is a fundamental challenge, and the inclusion of spin-orbit coupling increases the complexity. The Hohenberg-Kohn \cite{Hohenberg1964} and Kohn-Sham\cite{Kohn1965}  density  functional  theory  (DFT)  framework in the local-density and the generalized gradient approximations (GGA) completely fails to stabilize the magnetic moment on the iridium sites. Therefore, no first principles approach has been able to provide a handle on the key interactions, let alone the balance between the electron correlation and the spin-orbit coupling effects in determining the ground state of the system. In order to rationalize experimental observations, however, an assortment of "beyond DFT" methods, such as the DFT+$U$ \cite{Kim2008,Kim2017,Liu2015,He2015,He2015a} and various dynamical-mean-field-theory-based schemes\cite{Kotliar2005,Held2006,Park2008} have been employed on Sr$_{2}$IrO$_{4}$ \cite{arita2014mott,martins2011reduced,arita2012ab,zhang2013effective} involving fine tuning of both the on-site Hubbard $U$ parameter and the strength of spin-orbit coupling. Notably, Hubbard $U$ can be obtained from first-principles using, for example, the constrained random phase approximation  (cRPA) scheme, which then allows {\it ab initio} DFT+$U$ calculations. However, one still requires user intervention in the form of a judicious choice of local Wannier projections and subdivision of the single-particle Hilbert space, limiting the predictive power of the theory.\cite{aryasetiawan2004frequency,miyake2008screened,Liu2018}

Recent progress in constructing advanced density functionals offers a new pathway for addressing, at the first-principles level, the electronic structures of correlated materials. In particular, the strongly constrained and appropriately normed (SCAN) meta-GGA exchange-correlation functional\cite{Sun2015}\footnote{ 
We note that SCAN has a tendency to enhance magnetic moments in itinerant magnets, which reflects an oversensitivity of the so-called iso-orbital indicator ($\alpha$) used in SCAN to distinguish between various chemical bonding environments. Improvements in the SCAN functional in this connection\cite{furness2019enhancing,mejia2019analysis}, however, are not likely to significantly change the conclusions of the present study}, which obeys all known constraints applicable to a meta-GGA functional, has been shown to accurately predict many of the key properties of the undoped and doped La$_2$CuO$_4$ and YBa$_2$Cu$_3$O$_6$.\cite{Lane2018,Furness2018,zhang2020competing} In La$_2$CuO$_4$, SCAN correctly captures the magnetic moment in magnitude and orientation, the magnetic exchange-coupling parameter, and the magnetic form factor along with the electronic band gap, all in accord with the corresponding experimental values. In near-optimally doped YBa$_2$Cu$_3$O$_7$, using the SCAN functional, Ref. {\onlinecite{zhang2020competing}} identifies a landscape of 26 competing uniform and stripe phases. In Ref. {\onlinecite{zhang2020competing}}, the charge, spin, and lattice degrees of freedom are treated on an equal footing in a fully self-consistent manner for the first time to show how stable stripe phases can be obtained without invoking any free parameters. These results indicate that SCAN correctly captures many key features of the electronic and magnetic structures of the cuprates and, thus, provides a next-generation baseline for investigating the missing correlation effects, such as the quasiparticle lifetimes and waterfall effects. We note also that the transferability of SCAN to the wider class of transition-metal oxides has been demonstrated in Refs. \onlinecite{varignon2019mott,zhang2019symmetry}.

\begin{figure*}[t]
\includegraphics[scale=.27]{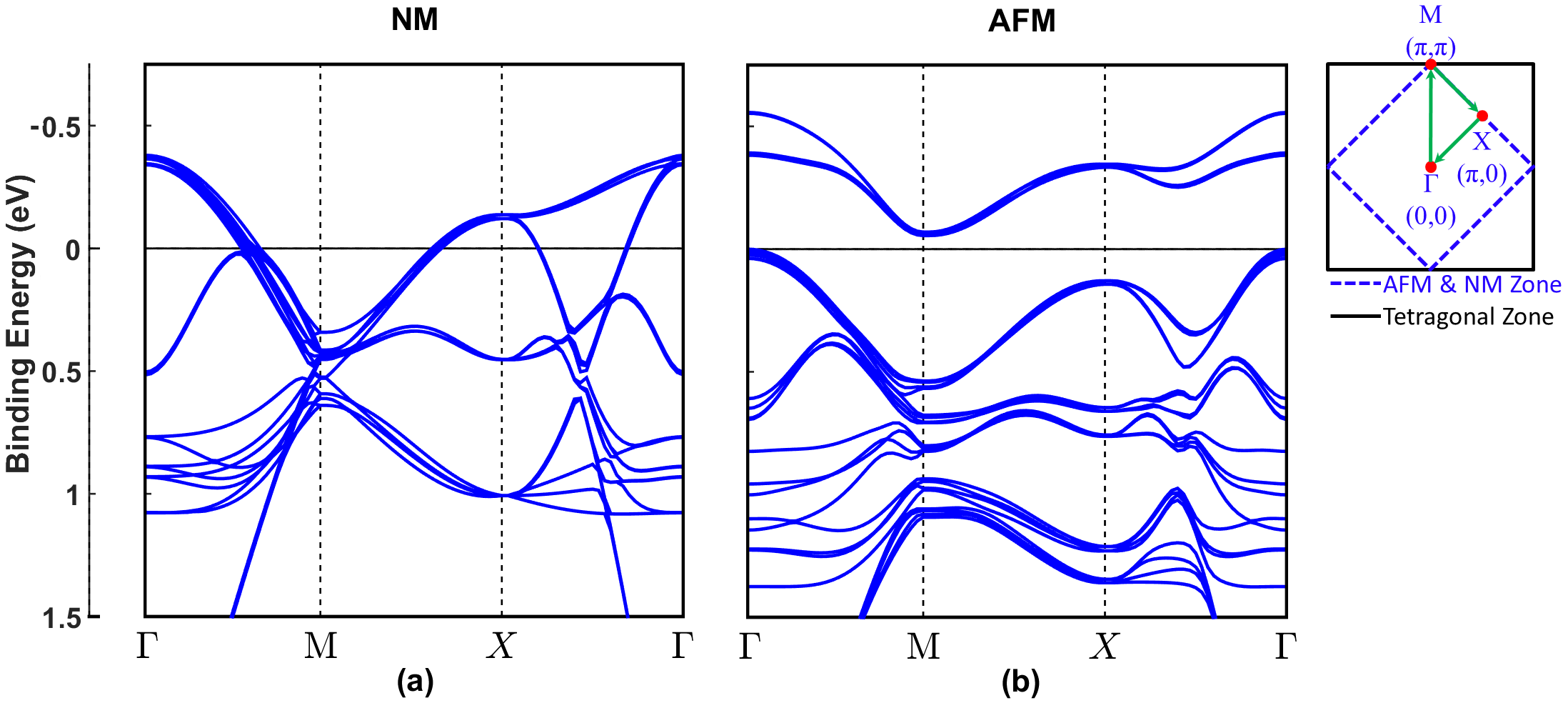}
\caption{(color online) (a) and (b) Electronic band dispersions of  Sr$_{2}$IrO$_{4}$ for the non-magnetic (NM) and antiferromagnetic (AFM) phases. A schematic of the AFM and NM Brillouin zones (blue dashed line) is shown on the right along with the tetragonal Brillouin zone (black solid line) for reference. The green arrows denote the high-symmetry lines along which the bands are plotted in panels (a) and (b).}
\label{fig:BANDS}
\end{figure*}

Here, we examine the efficacy of the SCAN functional in predicting the electronic and magnetic structures of Sr$_2$IrO$_4$. Our first-principles description of the magnetic ground state reproduces the key experimentally observed features of Sr$_2$IrO$_4$, including the size of the band gap and the magnitude and orientation of the Ir magnetic moments. By breaking the magnetic moments into their spin and orbital components, we show that the theoretically obtained magnetic scattering amplitude $\braket{M_{m}}$ as a function of polarization angle is consistent with NRMXS\cite{Fujiyama2014} measurements, indicating that SCAN correctly captures the delicate balance between the effects of electron correlations and spin-orbit coupling. Additionally, we predict appreciable magnetic moments on both the planar and the apical oxygen atoms. Finally, we compare Sr$_2$IrO$_4$ to the cuprates in terms of the so-called reference families\cite{markiewicz2017entropic} and show that Sr$_{2}$IrO$_{4}$ is similar to La$_{2}$CuO$_{4}$ and Bi$_{2}$Sr$_{2}$CaCuO$_{6}$.

\section{Computational Details}\label{sec:comp}
{\it Ab initio} calculations were performed using the pseudopotential projector-augmented wave method\cite{Kresse1999} implemented in the Vienna {\it ab initio} simulation package (VASP)\cite{Kresse1996,Kresse1993} with an energy cutoff of $650$ eV for the plane-wave basis set. Exchange-correlation effects were treated using the SCAN meta-GGA scheme\cite{Sun2015}. A 12 $\times$ 12 $\times$ 3 $\Gamma$-centered k-point mesh was used to sample the Brillouin zone. Spin-orbit coupling effects were included self-consistently. We used the experimental low-temperature I4/mmm crystal structure to initialize our computations. \cite{Huang1994} All atomic sites in the unit cell along with the cell dimensions were relaxed using a conjugate gradient algorithm to minimize the energy with an atomic force tolerance of $0.007$ eV/\AA~ and a total energy tolerance of $10^{-5}$ eV. Our theoretically obtained structural parameters are in accord with the corresponding experimental results. Our relaxed unit cell exhibits a slight 0.12\% orthorhombicity between the $a$- and the $b$-axes, consistent with the results of Porras {\it et al.} \cite{porras2019pseudospin}. As shown in Fig. \ref{fig:SPINSTRUCTURE}(a), the Sr$_2$IrO$_4$ structure can be viewed as a $\sqrt{2} \times  \sqrt{2}$ superlattice of I4/mmm symmetry in which alternating IrO$_{6}$ octahedra are rotated by 11.73$^{\circ}$. The rotational direction alternates within the layer as well as between the layers. In this way, the lattice can be subdivided into two sub lattices, and it can, therefore, intrinsically accommodate the AFM order without unit-cell doubling.\footnote{Note that all ordering vectors, $\mathbf{Q}$, are referenced with respect to the tetragonal I4/mmm structure, taking the staggered octahedral rotations as a distortion of the pristine phase. Therefore, the AFM order, originating on the iridium sites, is given by $\mathbf{Q}=(\pi,\pi)$, not the {na\"ively} assumed $\mathbf{Q}=(0,0)$ ordering vector.}

% Magnetic structure (spin structure)

\section{Magnetic Structure}\label{sec:magstruct}
The iridium magnetic moments in Sr$_2$IrO$_4$ are found  experimentally to be planar, following the staggered octahedral twists, producing a slight uncompensated ferromagnetic (FM) moment along the $a$- or $b$-axes. The IrO$_2$ layers are then stacked along the $c$-axis where the relative orientations of the FM moments produces six inequivalent magnetic configurations. \cite{porras2019pseudospin}  Our SCAN-based calculations show that the $+--+$ configuration of FM moments along the $b$-axis (as defined in Ref. \onlinecite{porras2019pseudospin}) is the ground state, with the remaining configurations lying $\sim 10^{-5}$ eV/Ir higher in energy. As expected, the lrrl arrangement is found to be equivalent to the $+--+$ stacking, except that the FM moments in the lrrl arrangement lie along the $a$ axis. The small energy separation of these states is consistent with estimates of the interlayer exchange energy\cite{Takayama2016} and suggests that these low-lying states would be  accessible to strong laser pump-probe spectroscopies.\cite{Zhao2015,DiMatteo2016}

Figure \ref{fig:SPINSTRUCTURE}(a) shows our theoretically obtained AFM state of Sr$_2$IrO$_4$ in the $+--+$ magnetic structure \cite{Takayama2016,Lovesey2012}. Two slightly inequivalent iridium magnetic sites are stabilized, labeled $A$ and $B$, as depicted in Fig.\ref{fig:SPINSTRUCTURE}(d). The predicted value of the magnetic moment is 0.237 $\mu_{B}$ and 0.250 $\mu_{B}$ on sites $A$ and $B$, respectively, in good accord with neutron diffraction studies\cite{Lovesey2012,Dhital2013,Ye2013,Ye2015}. Moreover, the iridium magnetic moment vector lies completely in the $ab$-plane, displaying a canted AFM ordering that follows the octahedral rotations. For the counter-clockwise (clockwise) twisted octahedra the magnetic moment is 2.9$^{\circ}$ (12.6$^{\circ}$) off of the $a$-axis. Due to magnetic moment canting, a slight uncompensated FM moment of 0.088 $\mu_{B}$ is produced directly 19.8$^{\circ}$ off of the $b$-axis in good accord with experimental studies.\cite{Crawford1994,Ge2011,Chen2015,Haskel2012,Takayama2016}

\begin{figure*}[t]
\includegraphics[scale=.29]{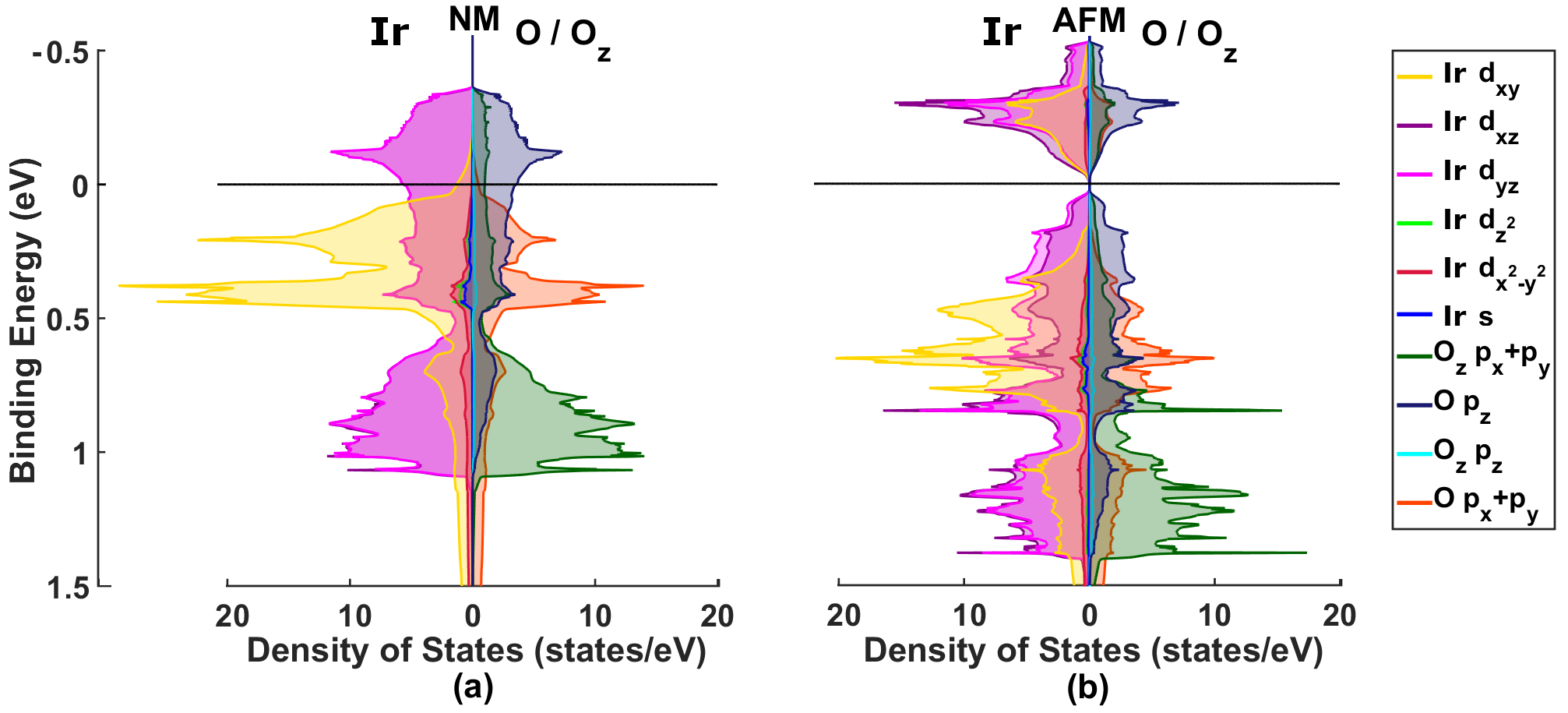}
\caption{(color online) Site-resolved partial densities-of-states in the NM and  AFM phases of Sr$_2$IrO$_4$. Iridium and oxygen characters are plotted on the left- and right-hand sides, respectively. Shadings and lines of various colors (see the legend) give contributions from various orbitals of  Ir, Oz, and in-plane O oxygen sites. }
\label{fig:DOS}
\end{figure*}

Figures \ref{fig:SPINSTRUCTURE}(b) and \ref{fig:SPINSTRUCTURE}(c) show the orbital and spin components of the total magnetic moment, shown in Fig. \ref{fig:SPINSTRUCTURE}(d), in a single IrO$_2$ plane. The spin component on site A (B) has a magnitude of 0.031 $\mu_{B}$ (0.044 $\mu_{B}$) with an angle off the a-axis of 45$^{\circ}$ (24$^{\circ}$). The orbital component is uniform across both Ir magnetic sites with a magnitude and orientation of 0.212 $\mu_{B}$ and 8.13$^{\circ}$, respectively. The inequivalence of sites A and B is likely due to effects of strong spin-lattice coupling, which breaks the four-fold rotational symmetry to produce the orthorhombic ground state.\cite{porras2019pseudospin} Moreover, when the crystal symmetry becomes lowered from tetragonal to orthorhombic, the intrinsic interlayer exchange coupling is no longer frustrated, so that it can contribute to the imbalance between the $A$ and $B$ sites.\cite{Takayama2016} A similar inequivalence between Ir sites has been reported in nonlinear optical harmonic generation measurements. \cite{Torchinsky2015}

Figure \ref{fig:SPINSTRUCTURE}(a) also displays magnetic moments on the planar oxygen atoms (dark blue and gold arrows). The predicted value of the magnetic moment on the oxygen atoms is 0.020 $\mu_{B}$ (blue) and 0.014 $\mu_{B}$ (gold) with the magnetic moment vector oriented completely in the $ab$-plane. The oxygen moments denoted by dark blue arrows are canted 11.31$^{\circ}$ off of the $a$-axis, whereas those denoted by gold arrows are only 4.08$^{\circ}$ off of the $a$-axis. As seen in Fig. \ref{fig:SPINSTRUCTURE}(b), the oxygen moments exhibit a purely spin character that form chains connecting the IrO$_6$ octahedra. Notably, recent first-principles calculations of La$_2$CuO$_4$ show a polarization of the in-plane oxygen $p_x$ and $p_y$ orbitals with no net magnetic moment\cite{Lane2018} because strong Cu-O hybridization and a Cu-O-Cu bond angle of 180$^{\circ}$ frustrate the oxygen magnetic density. Here, in contrast, oxygens carry a net moment driven by octahedral rotations, which break the magnetic density frustration on the oxygen sites. We find a small (0.008 $\mu_{B}$) apical-oxygen magnetic moment however, which is consistent with muon-spin-spectroscopy measurements. \cite{miyazaki2015evidence}

In order to determine the ground-state wave-function, estimates of the spin and orbital contribution to the ordered moment are necessary. Magnetic neutron scattering 
is usually employed to examine the local, microscopic magnetism in condensed matter systems. But neutrons cannot be used to separate orbital and spin contributions since neutrons do not interact with charges. The NRMXS technique, however, can probe both charge and magnetic degrees of freedom where the orbital and spin components can be separated via an analysis of the polarization dependence of the scattered x-rays.\cite{Blume1988,Grenier2014}  Fujiyama {\it et al.}\cite{Fujiyama2014} have reported the polarization angle ($\vartheta_{\text{pol}}$) dependence of the magnetic and charge scattering amplitude for three azimuthal angles from Sr$_2$IrO$_4$. By fitting the sinusoidal variation in the scattering amplitude to a simple model, where orbital ($L$) and spin ($S$) moments are considered collinear, they found a ratio of $\braket{L}/\braket{S}=5.0\pm 0.7$, which deviates markedly from the value of 4.0 expected for an ideal $J_{eff}=1/2$ system.

To test the validity of our first-principles modeling, we calculated the expected intensity of the magnetic scattering given by $I_{m}\propto \mu_{\pi^\prime}^{2}+\mu_{\sigma^\prime}^{2}$, where 
\begin{align}
\left( 
\begin{matrix}
\mu_{\sigma^\prime}\\
\mu_{\pi^\prime}
\end{matrix} 
\right)
&=
\braket{M_{m}}
\left( 
\begin{matrix}
\cos(\vartheta_{\text{pol}})\\
\sin(\vartheta_{\text{pol}})
\end{matrix} 
\right)
\end{align}
and the magnetic scattering amplitude $\braket{M_{m}}$ is
\begin{align}
\braket{M_{m}}&=-i\tau\left[ \mathbf{S}(\mathbf{K})\cdot \mathbf{B}+\mathbf{L}(\mathbf{K})\cdot \mathbf{B_{0}} \right].
\end{align}
Here, $\tau$ is the ratio of the incident photon energy and electron
rest mass, and $\mathbf{B}$ and $\mathbf{B_{0}}$ depend on the unit vectors of the propagation and the polarization of the incident and scattered x-rays, see Refs.\cite{Blume1988,Grenier2014} for details. By directly calculating $I_{m}$, we avoid difficulties in comparing theoretical and experimental branching ratios. Note that Fujiyama {\it et al.} neglected the finite angle between the spin and the orbital contributions in their fit. 

Figure \ref{fig:SPINSTRUCTURE}(e) compares the theoretically predicted magnetic scattering intensity with the experimental values from Ref.\onlinecite{Fujiyama2014} as a function of the polarization angle, $\vartheta_{\text{pol}}$, for various crystal orientations $\Psi$. Angle $\Psi=0$ is defined as the direction of the iridium magnetic moment. The magnetic scattering intensity for sites A and B are seen to follow the experimental values in good agreement. Utilizing the magnitude of our {\it ab initio} obtained magnetic moments on site A (B) we find a ratio $\braket{L}/\braket{S}$ of 13.64 (9.66). The enhancement of the ratio can be attributed to the noncollinearity of $L$ and $S$. Hence, SCAN correctly finds a larger deviation away from the conventional $J_{eff}=1/2$ description of the ground state.

\section{Electronic Structure}\label{sec:electronic}

\begin{figure}[h!]
\includegraphics[width=\columnwidth]{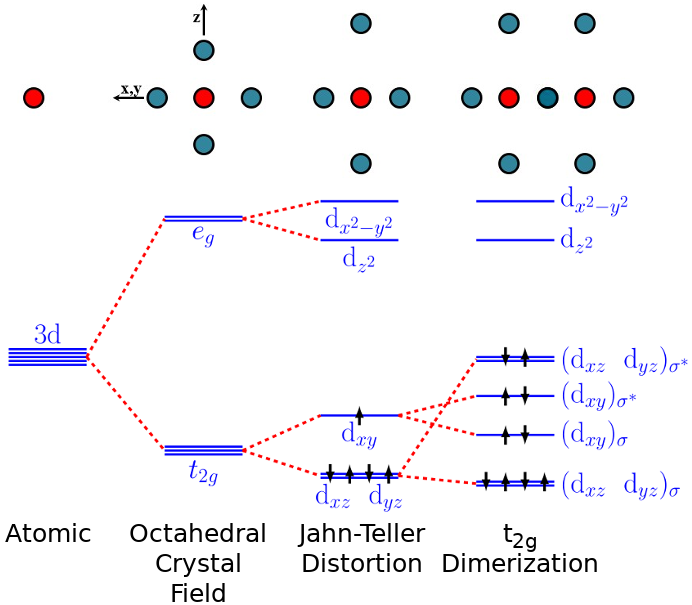}
\caption{(color online) Molecular orbital-energy level diagram for an isolated IrO$_{6}$ octahedron  under various crystal field conditions. A sketch of the atomic positions is given in the top portion of the figure, where Ir and O atoms are shown in red and blue colors, respectively. In an octahedral crystal field, the atomic Ir 5d levels split into $e_g$ and $t_{2g}$ manifolds. A positive tetragonal Jahn-Teller distortion splits the $e_g$ and $t_{2g}$ states as illustrated in the diagram. Introducing hybridization between the neighboring Ir sites produces bonding ($\sigma$) and antibonding ($\sigma^*$) states in the $t_{2g}$ manifold. Electron filling is indicated by black arrows, highlighting the difference in the highest occupied molecular orbital in the dimerized and non-dimerized cases.
}
\label{fig:MOLEBONDING}
\end{figure}

Figures \ref{fig:BANDS}(a) and \ref{fig:DOS}(a) show the band structure and partial density of states (DOSs) associated with various iridium and oxygen orbitals where the SOC strength is artificially set to zero. Here, and throughout, we will distinguish between the in-plane and the apical-oxygen atoms by O and Oz, respectively. Tuning SOC to zero, produces a metal in which the magnetic moment on all sites is zero. Here, the Fermi level cuts through the $t_{2g}$ states of the crystal-field split Ir 5d orbitals, with the $e_g$ states sitting 1.8 eV above. At the Fermi level, iridium-oxygen hybridized states dominate, where an atypical crystal field related stacking of states is seen. Specifically, the Ir d$_{xy}$ level is flanked above and below by the out-of-the-plane Ir d$_{xz}$ and d$_{yz}$ electrons, differing from the expected crystal-field split $t_{2g}$ states for an isolated elongated IrO$_6$ octahedron. This pattern is expected for iridium inter-site interactions facilitated by the delocalized nature of the 5d orbitals. The staggered octahedral rotations bend the Ir-O-Ir bonds, and thus enhance d$_{xz}$/d$_{yz}$ $\pi$-bonding while weakening the bonding between the adjacent d$_{xy}$ orbitals. The dimerization of neighboring iridium orbitals produces bonding and antibonding pairs of d$_{xz}$/d$_{yz}$ and alters the normal crystal-field ordering as shown in Fig.\ref{fig:MOLEBONDING}. This trend is consistent with the electron spin resonance study by Bogdanov {\it et al.} \cite{Bogdanov2015} and Gordon {\it et al.} \cite{Gordon2016}. 
Apical oxygen p$_{x}$ and p$_{y}$ orbitals also hybridize strongly with the Ir d$_{xz}$/d$_{yz}$ levels as pointed out by Agrestini {\it et al.}\cite{Agrestini2017}. There are small differences in band splitting at the $\Gamma$ point around 1  eV binding energy between our results and the LDA+$U$ calculations of Kim {\it et al}.\cite{Kim2008} This is due to our use of a relaxed orthorhombic structure whereas Kim {\it et al.} use a pristine tetragonal structure. The DOS and band structure over an extended energy range is given in Appendices \ref{A:fulldos} and  \ref{A:fullbands}, respectively.

Figures \ref{fig:BANDS}(b) and \ref{fig:DOS}(b) show the electronic band structure and DOS with the inclusion of SOC. The energy levels near the Fermi energy are now seen to reorganize, increasing the bandwidth of the Ir-d states due to SOC-induced splitting between the $J_{eff}=3/2$ and the $1/2$ bands. An AFM phase stabilizes with an optical gap at $M$ ($X$ and $\Gamma$)  of  592 meV (462 meV and 382 meV) in agreement with the $\alpha$ transition observed in optical conductivity studies, whereas the $\beta$ transition originates from valence bands $75$ meV below the Fermi energy. \cite{Kim2008,Moon2009,Propper2016,wang2018mott} Our transition energies are consistent with the state-of-the-art DFT+U$_{eff}$+SOC, BSE+GW+SOC\cite{Liu2018}, and DFT+DMFT\cite{arita2014mott,martins2011reduced,arita2012ab,zhang2013effective} calculations where the correlation strengths were estimated by constrained RPA, indicating that SCAN captures the subtle balance between the effects of SOC and electron correlations in Sr$_2$IrO$_4$. Our electronic structure is also in reasonable accord with with resonant inelastic x-ray scattering observations and a three-band tight-binding model fit to experimental optical conductivity and angle-resolved photoemission spectroscopy. \cite{wang2018mott,de2015collapse,ilakovac2019oxygen} Due to the indirect nature of the band gap, the gap in the DOS is only 55 meV, in agreement with electronic transport measurements \cite{Strydom2006,Zhou2017}. 

The opening of the band gap is typically ascribed to moderate on-site electron-electron interactions originating in a half-filled $J_{eff}=1/2$ band. A pure $J_{eff}=1/2$ state is composed of a linear combination of d$_{xz}$, d$_{yz}$, and d$_{xy}$ orbitals:
\begin{align}\label{Eq:basis1}
\ket{\psi_{+}}&=\frac{1}{\sqrt{N}}\left( iRd_{xy\downarrow}-\frac{1}{\sqrt{2}}d_{xz\uparrow}-\frac{i}{\sqrt{2}}d_{yz\uparrow} \right)\\
\ket{\psi_{-}}&=\frac{1}{\sqrt{N}}\left( -iRd_{xy\uparrow}+\frac{1}{\sqrt{2}}d_{xz\downarrow}-\frac{i}{\sqrt{2}}d_{yz\downarrow} \right)\label{Eq:basis2}
\end{align}
where the normalization factor $N$ and the relative weights of the $t_{2g}$-orbitals, $R$, are given in terms of the ratio ($\eta$) of the  strengths of the tetragonal crystal field and spin-orbit coupling:
\begin{align}\label{Eq:norm}
R(\eta)&=-\frac{1}{\sqrt{2}}\left( 1-\frac{1}{2}\left[ 1+2\eta+\sqrt{9-4\eta+4\eta^2} \right]  \right),\\
N(\eta)&=1+\frac{1}{2}\left( 1-\frac{1}{2}\left[ 1+2\eta+\sqrt{9-4\eta+4\eta^2} \right]  \right)^2,
\end{align}
as given in Ref. \onlinecite{DiMatteo2016}. For $\eta=0$, the ground state wave function is a pure $J_{eff}=1/2$ state.\cite{Kim2008} As the crystal-field splitting increases (decreases) the weight of the $d_{xy}$ state  increases (decreases).

Upon relaxing the crystal structure, we find a positive tetragonal distortion of 4.05\%. Our DOS, however, shows that the conduction and valence bands are dominated by d$_{xz}$/d$_{yz}$ orbital character with almost no d$_{xy}$ weight at the valence band edge. This is inconsistent with the pure $J_{eff}=1/2$ ground state, and more closely resembles the iridium inter-site hybridization scenario, where the bonding and anti-bonding bands reorganize the atomic character.\cite{Bogdanov2015} Due to the resulting dimerization, Eqs. (\ref{Eq:basis1}) and (\ref{Eq:basis2}) no longer describe the ground state. Strong O-$p_{z}$/Ir-d$_{xz/yz}$ hybridization is seen in Fig. \ref{fig:DOS}(b) with significant contributions to the valence and conduction states along with a nominal admixture of $e_{g}$ character. Taken together, these results show that the local $J_{eff}=1/2$ description of the low-energy electronic structure is modified via the non-local iridium inter-octahedra interactions. Consequently, the commonly employed one- and three-orbital tight-binding parametrizations of the iridates are of limited reach.

\section{Comparison with High-Temperature Cuprate Superconductors}
Like the cuprates, the low-energy physics in Sr$_{2}$IrO$_4$ is dominated by the single band that crosses the Fermi level. In order to compare the low-energy electronic structure of Sr$_{2}$IrO$_4$ with that of the cuprates, we follow the approach of Ref. \onlinecite{markiewicz2005} by constructing a one-band parametrization of the $J_{eff}=1/2$ state.
Here we emphasize that although the one-band model is useful due to its simplicity, the physical system involves the full manifold of Ir d-states and oxygen p-states. \footnote{ Note that for capturing the nesting features of the Fermi surfaces as a basis for discussing phase stabilities (via RPA susceptibilities), a one-band model is a reasonable starting point in both the iridates and the cuprates\cite{markiewicz2017entropic}). }

In this connection, we construct a single-band model in the pseudo-spin space with electron hopping between the Ir lattice sites as follows:
\begin{align}
H=\sum_{ij\sigma}t_{ij}c^{\dagger}_{i\sigma}c_{j\sigma}
\end{align}
where $c^{\dagger}_{i}(c_{j})$ create (destroy) fermions on the iridium site $i(j)$ with pseudospin eigenvalues $\sigma=\pm$. Since the iridium atoms sit on the vertices of a square lattice with hopping along the c-axis in a body-centered structure, the distance between the nearest neighbors, next-nearest neighbors, and so on, constrains the electronic dispersion allowing us to Fourier transform the Hamiltonian. The Hamiltonian can then be expressed as:
\begin{align}
H_{\mathbf{k}}=\sum_{\sigma}\left(\sum_{\braket{ij}}t_{ij}\exp{-i\mathbf{k}\cdot \mathbf{R}_{ij}}\right)c^{\dagger}_{\mathbf{k}\sigma}c_{\mathbf{k}\sigma}
\end{align}
with $\braket{ij}$ denoting that the sum is taken over successive rings of neighboring lattice sites surrounding site $i$, and $\mathbf{R}_{ij}$ is the displacement between the lattice sites $i$ and $j$. In general $H_{\mathbf{k}}$ can be rewritten as 
\begin{align}
H_{\mathbf{k}}=H^{\parallel}_{\mathbf{k_{\parallel}},k_{z}=0}+H^{\perp}_{\mathbf{k_{\parallel}},k_{z}}.
\end{align}
Here, $\mathbf{k_{\parallel}}$ and $k_{z}$ denote the in-plane and out-of-plane components of $\mathbf{k}$, respectively. We will now assume the interlayer coupling in the single Sr$_{2}$IrO$_4$ layer to be negligible, i.e. $H^{\perp}=0$, and only retain the dominant in-plane components. Taking the sum out to the fourth nearest neighbor, the dispersion is given by 
\begin{align}\label{eq:singelbandHAM}
&H_{k}=-2t(\cos(k_{x}a)+\cos(k_{y}a))\nonumber\\
&-4t^{\prime}(\cos(k_{x}a)\cos(k_{y}a))\nonumber\\
&-2t^{\prime\prime}(\cos(2k_{x}a)+\cos(2k_{y}a))\nonumber\\
&-4t^{\prime\prime\prime}(\cos(2k_{x}a)\cos(k_{y}a)+\cos(2k_{y}a)\cos(k_{x}a)),
\end{align}
where $a$ is the lattice spacing and the number of primes $(^{\prime})$ in the superscripts successively denote nearest neighbors, next-nearest neighbors, and so on.
We obtained the hopping parameters ($t$, $t^{\prime}$, ...) by fitting the one-band tight-binding dispersion to the first-principles band structure. The resulting hopping parameters are given in Table \ref{table:hopping1}. These parameters are quite similar to those adduced in Ref. \onlinecite{moutenet2018} obtained by down-folding a three orbital model.

\begin{table}[h]
\centering
\begin{tabular}{|c|c|c|c|}
$t$ & $t^{\prime}$ & $t^{\prime\prime}$ & $t^{\prime\prime\prime}$ \\
\hline
%207.0 &  49.8 & -20.0 &  -11.4\\
233.9  &  56.3 & -22.6 &  -12.9\\
\end{tabular}
\caption{Tight-binding hopping parameters (in meV) obtained by fitting our NM SCAN-based band structure.}\label{table:hopping1}
\end{table}

The strength of the effective on-site Hubbard interaction, $U_{eff}$, implied by our {\it ab initio} results, can be gauged by including a $(\pi,\pi)$ AFM order in our one-band model at the mean-field level (see appendix \ref{A:meanfield} for details). In this way, we find that for $U_{eff}=0.95$ eV, the one-band model can correctly match the value of the AFM gap in first-principles dispersion. This value of $U_{eff}$ is smaller by a factor of about $3$ compared to the typical value of $2.8$ eV in the cuprates\cite{mistark2015}, which is to be expected qualitatively since the relative bandwidth in Sr$_2$IrO$_4$ is smaller. The ratio $U_{eff}/t=4.06$ is approximately a factor of $2$ smaller than for various cuprates\cite{mistark2015}, which would place Sr$_2$IrO$_4$ firmly in the middle of the mean-field phase diagram of the Hubbard model\cite{claveau2014mean}.

In Ref.~\onlinecite{markiewicz2017entropic}, Markiewicz {\it et al}. introduce `reference families' for characterizing different classes of materials. In order to reduce the number of parameters needed to describe various types of first-principles electronic dispersions, the idea is to map the system onto a small set of standardized hopping parameters or the so-called {\it reference families}. Equivalence between various model Hamiltonians is then determined by examining their `fluctuation  phase diagrams' which refer to the map of the leading instability of the system as given by the Stoner criteria, $1-U\chi_{0}(\omega=0,q)=0$, as a function of doping and temperature; see Ref. \onlinecite{markiewicz2017entropic} for details. Specifically, Hamiltonians with similar fluctuation phase diagrams are classified as being equivalent. Once a material system is mapped into a reference family, one can compare and contrast its properties with other down-folded materials (e.g., the cuprates), to help search for new materials with similar properties.

Comparing the effective hopping parameters in Table \ref{table:hopping1} with those of the high-temperature cuprate superconductors in Fig. 5 of Ref. \onlinecite{markiewicz2017entropic}, we adduce that Sr$_{2}$IrO$_{4}$ sits on the boundary between the La$_{2}$CuO$_{4}$ and the Bi$_{2}$Sr$_{2}$CaCuO$_{6}$  reference families.
\footnote{Note that in order to compare the electron-doped Sr$_2$IrO$_4$ and the hole-doped cuprates, we must perform an electron-hole transformation on the Sr$_2$IrO$_4$ dispersion by changing the signs of $t^{\prime}$ and $t^{\prime\prime}$.}
This conclusion is in keeping with various experimental results\cite{squarelatticeiridate} which show that the single layer iridates follow a wide range of cuprate phenomenology, suggesting that the iridates should exhibit superconductivity. In fact, Sr$_{2}$IrO$_{4}$ seems to fall in a very interesting parameter range between La$_{2}$CuO$_{4}$ and most other cuprates and may potentially lie close to the Mott-Slater crossover.\cite{markiewicz2017entropic} Also, the proximity of Sr$_{2}$IrO$_{4}$ to Bi$_{2}$Sr$_{2}$CaCuO$_{6}$ suggests the possible existence of incommensurate charge-density-wave phases in the iridates, which have been recently observed in scanning-tunneling measurements\cite{wang2019doping}.

%% Conclusions

\section{Summary and Conclusion}\label{sec:conclusion}

We have demonstrated that a first-principles treatment of the magnetic structure of the AFM ground state of Sr$_2$IrO$_4$ is possible without invoking any free parameters, and thus capture correctly the delicate balance between the effects of spin-orbit coupling and electron-electron correlations. We show that iridium inter site interactions play an important role in the electronic structure, so that local, one-band low-energy effective models are intrinsically of limited reach in 5d electron systems. Our treatment will be of value more generally for parameter-free examination of electronic structures, magnetism, and phase diagrams of other spin-orbit driven correlated materials.

%% Acknowledgments

\begin{acknowledgments}

This work was supported (testing efficacy of new functionals in complex materials) by the U.S. Department of Energy (DOE) Energy Frontier Research Centers: Center for the Computational Design of Functional Layered Materials (Grant No. DE-SC0012575). The work at Northeastern University was also supported by the DOE, Office of Science, Basic Energy Sciences (Grant No. DE-FG02-07ER46352) (core research) and benefited from Northeastern University's Advanced Scientific Computation Center, the National Energy Research Scientific Computing Center Supercomputing Center (DOE Grant No. DE-AC02-05CH11231). The work at Tulane was also supported by the DOE under EPSCoR Grant No. DE-SC0012432 with additional support from the Louisiana Board of Regents. B.B. acknowledges support from the COST Action CA16218.

\end{acknowledgments}

\appendix

\section{Extended Site-Projected Density of States}\label{A:fulldos}
\begin{figure*}[h!]
\includegraphics[width=.8\textwidth]{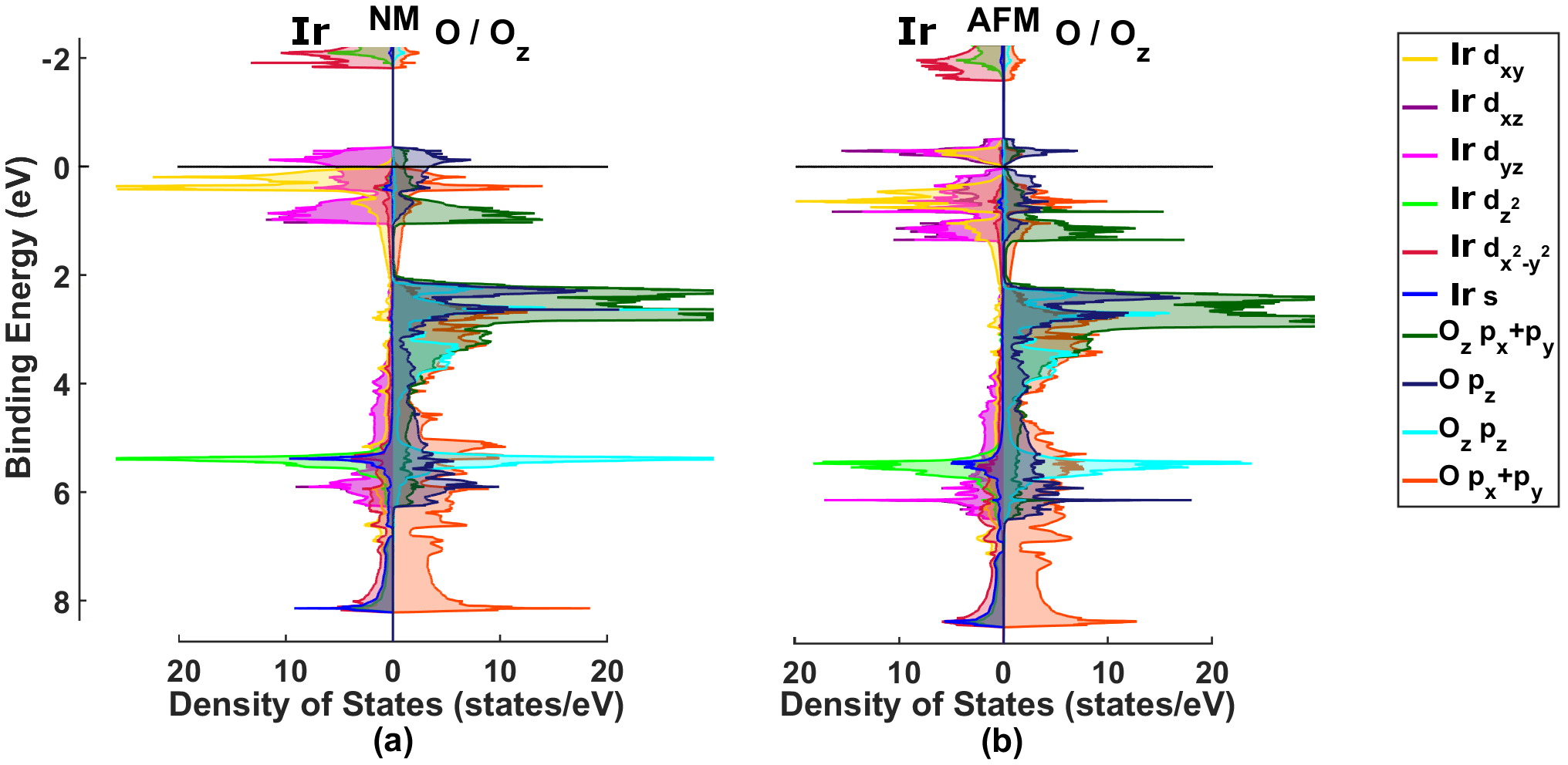}
\caption{(color online) Same as Fig. \ref{fig:DOS}, except that this figure shows the full electronic Ir-$d$ bandwidth.  }
\label{fig:EXDOS}
\end{figure*}

Figure \ref{fig:EXDOS} shows the site-resolved partial density-of-states for various iridium 5d and oxygen 3p orbitals over an extended binding energy range covering the full bandwidth with and without spin-orbit coupling. Figure \ref{fig:EXDOS}a shows a "mirroring" of $e_g$ states at larger binding energies similar to the cuprates\cite{Lane2018}. The oxygen-hybridized $e_g$ states form the bottom of the band, where the Ir d$_{x^2-y^2}$ states dominate, producing a clear one-dimensional van Hove singularity at the band edge. Moreover, Ir d$_{z^2}$ / O$_z$ p$_{z}$ hybridization produces an intense van Hove singularity at 5.5 eV binding energy and Ir d$_{xz}$(d$_{yz}$) / O p$_{z}$ states dominate around 6 eV binding energies. In contrast to the cuprates, strong O$_z$ p$_{x}$(p$_{y}$)/ O p$_{z}$ bonding is found at intermediate binding energies of 2.5 eV. Figure \ref{fig:EXDOS}b exhibits the same features below 2 eV as Fig. \ref{fig:EXDOS}a, showing that spin-orbit coupling has little effect on these states.

\section{Extended Band Structure}\label{A:fullbands}
\begin{figure*}[h!]
\includegraphics[width=.9\textwidth]{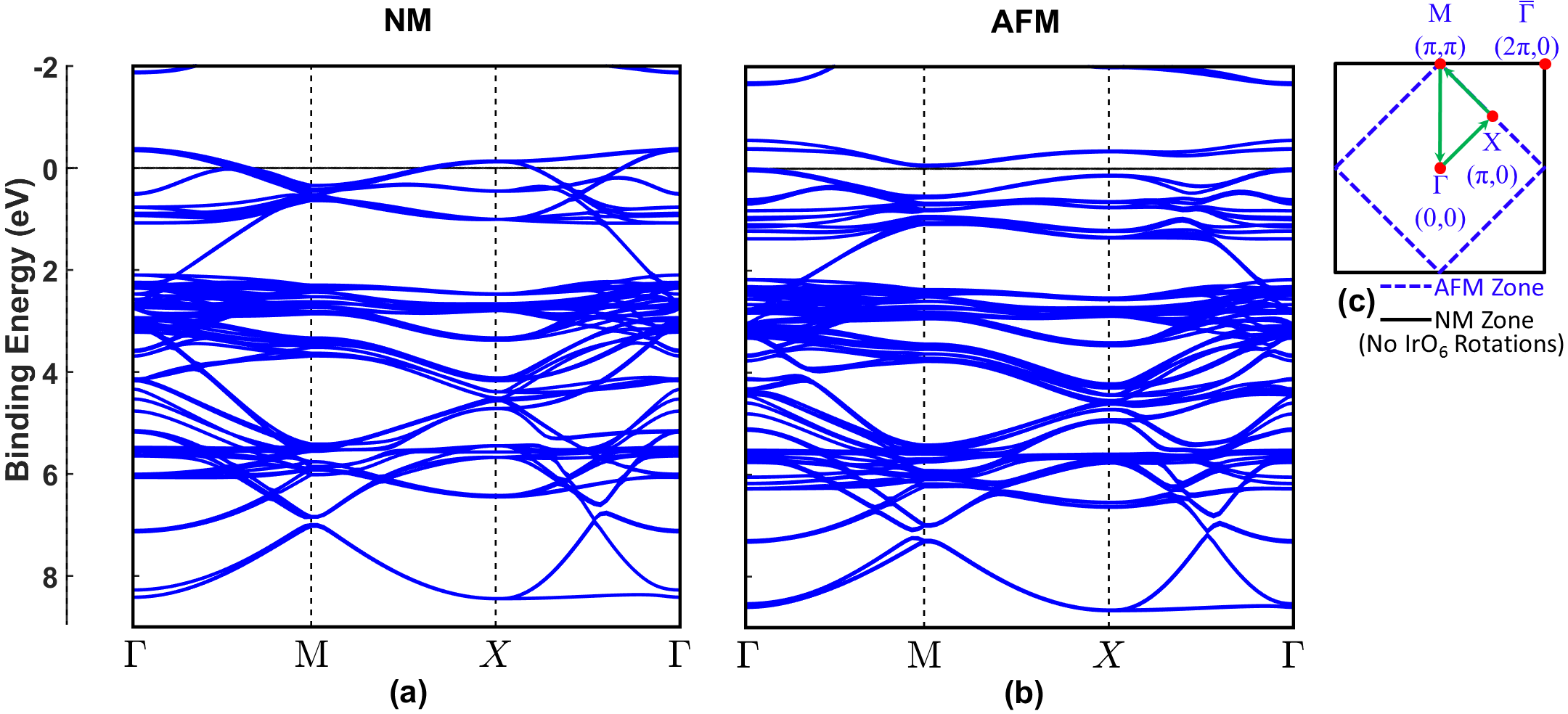}
\caption{(color online) Same as Fig. \ref{fig:BANDS}, except for an extended energy window covering the full Ir-d bandwidth in the NM and AFM phases }
\label{fig:EXBANDS}
\end{figure*}

Figure \ref{fig:EXBANDS} shows the band structure (blue lines) of Sr$_2$IrO$_4$ over an extended energy window covering the full bandwidth in the NM and AFM phases.

\section{Mean-Field Interactions and $(\pi,\pi)$-AFM Order}\label{A:meanfield}

In order to account for the staggered AFM order on the iridium sites, we include an on-site Hubbard interaction term to the Hamiltonian of Eq.  (\ref{eq:singelbandHAM}). Specifically, the double-occupancy energy penalty $U$ is placed on the single effective band crossing the Fermi level. The Hubbard interaction can be written in momentum space as
\begin{align}
\frac{U}{2}\sum_{\sigma}\sum_{kk^{\prime}Q} c^{\dagger}_{k\sigma} c_{k\sigma} c^{\dagger}_{k^{\prime}\bar{\sigma}} c_{k^{\prime}\bar{\sigma}}+c^{\dagger}_{k+Q\sigma} c_{k\sigma} c^{\dagger}_{k^{\prime}\bar{\sigma}} c_{k^{\prime}+Q\bar{\sigma}}.
\end{align}
where $\bar{\sigma}$ denotes $-\sigma$. Due to momentum conservation, the interaction depends on both the crystal momentum, $k(k^{\prime})$, of the electrons and the momentum transferred, $Q$, during the interaction. The momentum transfer gives rise to umklapp processes where electrons can scatter to neighboring Brillouin zones, which are the key for describing various density-wave instabilities. Here we take $Q=(\pi,\pi)$ following the experimentally observed AFM order. Thus, the full single-band Hamiltonian is  
 \begin{align}\label{Eq:Hamint}
\mathcal{H}&=
\sum_{\sigma}\sum_{k} \left( H_{k\sigma} c^{\dagger}_{k\sigma}c_{k\sigma} + H_{k+Q\sigma} c^{\dagger}_{k+Q\sigma}c_{k+Q\sigma} \right)\nonumber\\
&-\mu\sum_{\sigma}\sum_{k}\left(\hat{n}_{k\sigma}+\hat{n}_{k+Q\sigma} \right)\nonumber\\
&+\frac{U}{2}\sum_{\sigma}\sum_{kk^{\prime}} c^{\dagger}_{k\sigma} c_{k\sigma} c^{\dagger}_{k^{\prime}\bar{\sigma}} c_{k^{\prime}\bar{\sigma}}+c^{\dagger}_{k+Q\sigma} c_{k\sigma} c^{\dagger}_{k^{\prime}\bar{\sigma}} c_{k^{\prime}+Q\bar{\sigma}}.
\end{align}
where $H_{k}$ is written in terms of $Q$ explicitly by restricting $k(k^{\prime})$ to the smaller AFM Brillouin zone. We now rewrite the interaction in terms of the mean-field and expand the number operator in terms of fluctuations away from the mean electron count per state $\braket{n_{k\sigma}}$,
 \begin{align}
n_{k\sigma} &= \braket{n_{k\sigma}}+\left( n_{k\sigma}-\braket{n_{k\sigma}} \right)\\
&= \braket{n_{k\sigma}}+\delta_{\sigma},\nonumber
\end{align}
where $\delta_{\sigma}$ is the fluctuation away from $\braket{n_{k\sigma}}$. We substitute into the interaction of Eq. (\ref{Eq:Hamint}) assuming fluctuations are small $\delta_{\sigma}\delta_{\bar{\sigma}}\approx 0$ giving
\begin{align}\label{eq:meanmat}
\frac{U}{2}\sum_{\sigma}\sum_{kk^{\prime}} \braket{c^{\dagger}_{k\sigma} c_{k\sigma}}  c^{\dagger}_{k^{\prime}\bar{\sigma}} c_{k^{\prime}\bar{\sigma}}+\braket{c^{\dagger}_{k^{\prime}\bar{\sigma}} c_{k^{\prime}\bar{\sigma}}} c^{\dagger}_{k\sigma} c_{k\sigma}\nonumber\\
+\braket{c^{\dagger}_{k+Q\sigma} c_{k\sigma}} c^{\dagger}_{k^{\prime}\bar{\sigma}} c_{k^{\prime}+Q\bar{\sigma}}\nonumber+\braket{c^{\dagger}_{k^{\prime}\bar{\sigma}} c_{k^{\prime}+Q\bar{\sigma}}} c^{\dagger}_{k+Q\sigma} c_{k\sigma}\nonumber.
\end{align}
In order to treat the various matrix elements in Eq. (\ref{eq:meanmat}), we consider the average charge and spin densities as a function of momentum transfer $q$,
\begin{align}
\braket{\rho(q)}&=\sum_{k}\braket{\left(c_{k+q\uparrow}^{\dagger} c_{k+q\downarrow}^{\dagger} \right) \mathbb{I}  \left( \begin{array}{c} c_{k\uparrow} \\ c_{k\downarrow} \end{array} \right)}\\
&=\sum_{k}\braket{c_{k+q\uparrow}^{\dagger}  c_{k\uparrow}}+\braket{c_{k+q\downarrow}^{\dagger}  c_{k\downarrow}}\nonumber\\
&=N_{e}\delta_{q,0}\nonumber\\
\braket{S^{z}(q)}&=\frac{1}{2}\sum_{k}\braket{\left(c_{k+q\uparrow}^{\dagger} c_{k+q\downarrow}^{\dagger} \right) \sigma^{z}  \left( \begin{array}{c} c_{k\uparrow} \\ c_{k\downarrow} \end{array} \right)}\\
&=\frac{1}{2}\sum_{k} \braket{c_{k+q\uparrow}^{\dagger}c_{k\uparrow}}-\braket{c_{k+q\downarrow}^{\dagger}c_{k\downarrow}}.  \nonumber
\end{align}
Therefore, for $q=Q=(\pi,\pi)$,
\begin{align}
\braket{\rho(Q)}&=\sum_{k}\braket{c_{k+Q\uparrow}^{\dagger}  c_{k\uparrow}}+\braket{c_{k+Q\downarrow}^{\dagger}  c_{k\downarrow}}\\
&=0\nonumber
\end{align}
which implies
\begin{align}\label{eq:homoresult}
\braket{c_{k+Q\uparrow}^{\dagger}  c_{k\uparrow}}=-\braket{c_{k+Q\downarrow}^{\dagger}  c_{k\downarrow}}.
\end{align}
Also, by Hermiticity we have the equivalence, 
\begin{align}
\braket{c_{k+Q\sigma}^{\dagger}  c_{k\sigma}}^{\dagger}=\braket{c_{k\sigma}^{\dagger}  c_{k+Q\sigma} }.
\end{align}
Using the relation in Eq.\ref{eq:homoresult} we find $\braket{S^{z}(Q)}$,
\begin{align}
\braket{S^{z}(Q)}&=\frac{1}{2}\sum_{k} \braket{c_{k+Q\uparrow}^{\dagger}c_{k\uparrow}}-\braket{c_{k+Q\downarrow}^{\dagger}c_{k\downarrow}}\\
&=\sum_{k} \braket{c_{k+Q\uparrow}^{\dagger}c_{k\uparrow}}.\nonumber
\end{align}
The preceding relations allow us to cast staggered magnetization and electron density as
\begin{align}
m&=\sum_{k} \braket{c_{k+Q\uparrow}^{\dagger}c_{k\uparrow}}=-\sum_{k} \braket{c_{k+Q\downarrow}^{\dagger}c_{k\downarrow}},\nonumber\\
n_{\sigma}&=\sum_{k} \braket{c_{k\sigma}^{\dagger}c_{k\sigma}}.
\end{align}
Inserting these definitions and simplifying we arrive at the Hamiltonian in terms of the self-consistent field $m$ and occupation $n_{\sigma}$,
\begin{align}\label{eq:scham}
H_{k\sigma}&=\begin{bmatrix}
    H_{k\sigma}+Un_{\bar{\sigma}}&sign(\bar{\sigma})\Delta\\
    sign(\bar{\sigma})\Delta&  H_{k+Q\sigma}+Un_{\bar{\sigma}}  \\
\end{bmatrix}
\end{align}
where our wave functions take the Nambu form $\Psi=\left( c^{\dagger}_{k\sigma}~,~ c^{\dagger}_{k+Q\sigma} \right)$ and $\Delta$ is defined as $\frac{U}{2}\left(m+m^{\dagger}\right)=URe\left(m\right)$.

To self consist $m$ and $n$, their expectation value can be written in terms of the diagonalized system. Let the quasiparticle creation $(\gamma_{k\mu}^{\dagger})$ and annihilation $(\gamma_{k\mu})$, operators in the diagonalized system be defined as
\begin{align}
c_{k\sigma}&=\sum_{\mu}V^{k}_{\sigma,\mu}\gamma_{k\mu} ~~\text{and}~~ c_{k\sigma}^{\dagger}=\sum_{\mu}\gamma^{\dagger}_{k\mu}(V^{k}_{\sigma,\mu})^{\dagger}.
\end{align}
where $\mu$ indexes the bands. Therefore $m$ and $n$ are given by
\begin{align}
n_{\sigma}&=\sum_{\mu}\sum_{k} \left( (V^{k}_{\sigma\mu})^{\dagger}    V^{k}_{ \sigma \mu} + (V^{k+Q}_{\sigma \mu})^{\dagger}    V^{k+Q}_{\sigma \mu}  \right) 
f(\epsilon_{k\sigma \mu}),\\
m&=\sum_{\mu}\sum_{k} \left( (V^{k+Q}_{\sigma \mu})^{\dagger}    V^{k}_{ \sigma \mu} + (V^{k+Q}_{\sigma \mu})^{\dagger}    V^{k+Q}_{\sigma \mu}  \right) 
f(\epsilon_{k\sigma \mu}).
\end{align}
for $k$ in the AFM Brillouin zone and $f$ being the Fermi function. The self-consistently obtained values of the expectation value of $m$ and $n_{\sigma}$ are 0.33379 and 0.49729, respectively, within a tolerance of $10^{-5}$ at a temperature of $0.001$ K. 

*Corresponding authors: Christopher Lane (c.lane@neu.edu), Jianwei Sun (jsun@tulane.edu), Arun Bansil (ar.bansil@neu.edu)

\twocolumngrid

\bibliography{Iridate_Refs}

\end{document}